\documentclass[conference]{IEEEtran}
\usepackage{balance}

\IEEEoverridecommandlockouts 
\usepackage[dvips]{graphicx}
\usepackage{algorithmic}
\usepackage{algorithm}
\usepackage{listings}
\usepackage[OT4,T1]{fontenc}
\usepackage[cmex10]{amsmath}
\usepackage{url}
\usepackage{multirow}

\title{Partner selection supports reputation-based cooperation in a Public Goods Game}

\author{
\IEEEauthorblockN{Daniele Vilone, Francesca Giardini and Mario Paolucci}
\IEEEauthorblockA{
\ \\
LABSS (Laboratory of Agent Based Social Simulation)\\
Institute of Cognitive Sciences and Technologies, National Research Council (CNR)\\
via Palestro 32, 00185 Rome, Italy\\
\ \\
Email: daniele.vilone@gmail.com}
}

\begin{document}
\maketitle   
\begin{abstract}

In dyadic models of indirect reciprocity, the receivers' history of giving has a significant impact on the donor's decision. When the interaction involves more than two agents things become more complicated, and in large groups cooperation can hardly emerge. In this work we use a Public Goods Game to investigate whether publicly available reputation scores may support the evolution of cooperation and whether this is affected by the kind of network structure adopted. Moreover, if agents interact on a bipartite graph with partner selection cooperation can thrive in large groups and in a small amount of time.
\end{abstract}

\ 

KEYWORDS: Evolution of cooperation; Public Goods Game; Network; Reputation.

\ 

\section{Introduction}

\ 

{\it
Two neighbors may agree to drain a meadow, which they possess in common: because it is easy for them to know each other's mind;
and each must perceive, that the immediate consequence of his failing in his part, is the abandoning of the whole project.
But it is very difficult, and indeed impossible, that a thousand persons should agree in any such action; it being difficult
for them to concert so complicated a design, and still more difficult for them to execute it; while each seeks a pretext to free
himself of the trouble and expense, and would lay the whole burden on the others.
}

\ 

David Hume, A Treatise of Human Nature (London: J. M. Dent, 1952, II, p.239) - From reference~\cite{car07}.

\ 

Humans show levels of cooperation among non-kin that are unparalleled among other species. This difference becomes striking when facing social dilemmas, i.e., situations in which cooperation is hard to achieve because the best move for an individual does not produce the best outcome for the group. Public goods games (PGG) represent a clear exemplification of this conflict between individual incentives and social welfare. If everybody contributes to the public good, cooperation is the social optimum, but free-riding on others' contributions represent the most rewarding option. 

If norms, conventions and societal regulations have been proven effective in preventing the collapse of public goods (for a review, see~\cite{ost90,ost05}), when individuals are faced with unknown strangers, with little or no opportunities for future re-encounters cooperation easily collapses, unless punishment for non-cooperators is provided~\cite{feh00}. 
An alternative solution is represented by reputation, through which cheaters can be easily identified and avoided~\cite{now98a,gia12}. Indirect reciprocity supported by reputation~\cite{ale87} can be one of the mechanisms explaining the evolution of cooperation in humans~\cite{mil02}, especially in large groups of unrelated strangers who can, through language, actively communicate about their past experiences with cheaters~\cite{smi10}.

As such, gossip may effectively bypass the ``second-order free-rider problem'', wherein the costs associated with solving one social dilemma produces a new one~\cite{har68,kiy08}. This is the case of punishment: cooperators who do not sustain the costs of punishment are better off than cooperators who also punish. Therefore this solution to social dilemmas itself entails a social dilemma, whereas gossip, being essentially free should not imply such a second-order free-rider problem. In addition to costly punishment and reputation, ostracism of free-riders may represent a third solution. However, the direct effect of ostracizing a member is that the group size decreases, which automatically reduces maximal contribution levels to the public good for all remaining periods. Maier-Rigaud and colleagues show that in the lab PGG with ostracism opportunities increases contribution levels and contrary to monetary punishment, also has a significant positive effect on net earnings~\cite{mai05}.

Models of indirect reciprocity usually take into account dyadic interactions~\cite{now98a}, or group interactions in a mutual aid game~\cite{pan04}, in which providing help has a cost for the helper but it also increases his/her image score, i.e., a publicly visible record of his/her reputation. Image score increases or decreases according to individuals' past behaviors, thus providing a reliable way to discriminate between cheaters and cooperative players. Both in computer simulations~\cite{now98a}, and in lab experiments with humans~\cite{wed00}, cooperation can emerge and be maintained through image score.

When individuals facing a social dilemma can know other players' image score, cooperation can emerge in small groups, as showed by Suzuki and Akiyama~\cite{suz05}. In their work, cooperation can emerge and be maintained for groups of four individuals; though, when group size increases there is a concomitant decrease in the frequency of cooperation. The authors explain this decline as due to the difficulty of observing reputations of many individuals in large communities. This can be true of unstructured communities, but this rarely happens in human societies, characterized by interaction networks. To account for the role of societal structure, we designed a PGG in which players' interactions depend on the kind of network and on the possibility of actively choosing a subset of group members. More specifically, we compare cooperation levels among agents placed on a small-world network~\cite{wat98}, defined by short average path lengths and high clustering, to the performance of agents on a bi-partite graph~\cite{die97,gar11}. The latter is generally used to model relations between two different classes of objects, like affiliation networks linking members and the groups they belong to. This structure is especially interesting for us because it is especially suited for partner selection, as it happens when a club refuses membership to a potential associate.

Here, we are interested in exploring the effect of network structure on the emergence of cooperation in a PGG. We compare two different network topologies and we show that reputation-based partner choice on a bi-partite graph can make cooperation thrive also in large groups of agents. We also show that this effect is robust to number of generations, group size and total number of agents in the system. 

\ 

\ 

\section{The Model}

We consider a population of $N$ individuals. In each round of the game, $g$ agents are picked up at random to play a PGG among themselves.
Players can cooperate contributing with a cost $c$ to a common pot, or can defect without paying anything. Then, the total amount collected
in the pot is multiplied for a benefit $b$ and equally distributed among all the group members, without taking into account individual
contributions. At the end of each interaction, said $\chi$ the number of contributors in the group, cooperators' payoffs equals $(\chi b/g-1)c$,
whereas defectors' payoffs is $\chi bc/g$. At the collective level, the best outcome is achieved when everyone cooperates, but cheaters are better
off, because defection permits to avoid a loss when the number of cooperators is lower than $gc/b$.

Among the many solutions offered~\cite{feh00}, Suzuki and Akiyama~\cite{suz05} design a modified PGG in which agents can identify cheaters thanks to the so-called image score ~\cite{now98a,now98b}. The basic features of our model are the same of the one by Suzuki and Akiyama: in particular, each player $i$ is characterized by two integer variables:
the image score $s_i\in[-S_{max},S_{max}]$ and
the strategy $k_i\in[-S_{max},S_{max}+1]$, being $S_{max}\geq0$ a parameter of the model. When selected to play a round of the game, an individual
cooperates if the average image score $\langle s\rangle_g$ of its opponents is equal to or higher than its own strategy $k_i$, otherwise
it will not contribute. At the end of the round, the image score of the player is increased by 1 in case of cooperation, otherwise it is
decreased by the same quantity. In any case, $s_i$ remains in the allowed interval $[-S_{max},S_{max}]$: if an agent has an image score of
$S_{max}$ ($-S_{max}$) and contributes (defects), nothing happens to its image score. At the initial stage, all the
image scores and fitness levels are set to zero, whilst the strategies are randomly distributed among the individuals.

The image score is intended to give a quantitative evaluation of the public reputation of an individual in the scope of indirect reciprocity:
if contributing once is rewarded by future contributions by the other individuals, then any cooperative action must be
recognized and considered positively by the entire population; on the other hand, the variability of the strategies describes the different
attitudes and expectations of the single agents~\cite{now98a}.

After $m$ rounds, reproduction takes place. Again, we apply the same evolutionary algorithm used by Suzuki and Akiyama~\cite{suz05}. For $N$ times we
select at random a pair of individuals and with probability $P$ we create a new individual inheriting the strategy of the parent with the highest 
fitness. Then parents are put again in the population, and offspring is stored in another pool. When this selection process
has happened $N$ times, the old population is deleted and replaced with the offspring. It is worth noticing that offspring
inherit only the parent's strategy, while their image score and fitness is set equal to zero. Finally, we repeat all the procedure ($m$
rounds followed by the reproduction stage), for an adequate number of generations. The simulation lasts until the system reaches a final
(steady or frozen) configuration.

For sake of clarity, we observe that strategies defined as ($k\leq0$) are the more ``cooperation prone'', with the limit
case of $k=-S_{max}$ which is an absolute cooperator, while the positive ones are the ``cooperation averse'' strategies, with the limit case
of $k=S_{max}+1$ representing an inflexible defector.

Moving from the model described above, we are interested in testing whether two different network structures can promote cooperation for different group size and what effect partner selection can have in such an environment. 

\ 

\ 

\section{Results}

\ 

\subsection{Robustness of Suzuki's and Akiyama's results}

Suzuki and Akiyama tested their model for a given set of parameters with the following values: $N=200,\ c=1,\ b=0.85g,\ S_{max}=5,
\ m=800$. Their results show that  a cooperative strategy can evolve and invade a population when group size $g$ is small, but it does not survive when groups are large. For medium-sized communities, a coexistence between cooperators and defectors is possible.

\ 

The first step of the study present in this paper is a check of the robustness of Suzuki and Akiyama results with respect to the values of the
model parameters. A check of the role of $m$ and $N$ is reported already in~\cite{suz05}: it is claimed that the outcome is not relevantly
influenced by the value of these two quantities, so we focus here on $b,\ P$ and $S_{max}$.

\begin{figure}[tbp]
\centering
\includegraphics*[width=0.5\hsize]{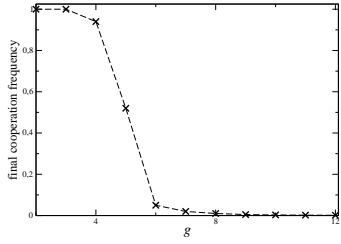}
\caption{Behaviour of the final frequency of cooperative actions as a function of the group size $g$. All the parameters are the same of
reference~\cite{suz05}, except $b=3$. Each point averaged over 1000 realizations.}
\label{B3_check}
\end{figure}

The role of $b$ in the PGG is quite clear in literature. Normally it is set to a fixed value larger than one (often 3), independent
from the group size~\cite{hau02}. Using this value,
we found that the final cooperation level decreases sharpenly as $g$ increases, as shown in Fig.~\ref{B3_check}. The fact that in Suzuki's
and Akiyama's work such decreasing is much slower is due to the fact that being $b$ proportional to the group size the number of contributors
needed in order to make cooperation convenient remains constant in $g$ instead of decreasing with it. On the other hand, even though less dramatic,
the decrease is anyway observed, indicating that the negative effect of large groups on cooperation is stronger and it might depend on the PGG
dynamics itself.

\begin{figure}[tbp]
\centering
\includegraphics*[width=0.5\hsize]{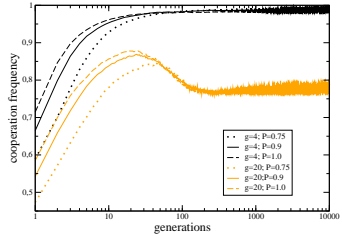}
\caption{Behaviour of the frequency of cooperative actions as a function of the number of generations for three different
values of $P$: 0.75, 0.90 and 1.0. The remaining parameters are the same of reference~\cite{suz05}. Each curve averaged over 1000 realizations.}
\label{P_check}
\end{figure}

Concerning the behaviour of the model as a function of the parameter $P$, we tested three different values: $P=0.9$ as in~\cite{suz05},
$P=0.75$ and $P=1.0$. As it can be easily seen in Fig.~\ref{P_check}, there is no fundamental difference due to the exact value of
this parameter.

\begin{figure}[tbp]
\centering
\includegraphics*[width=0.5\hsize]{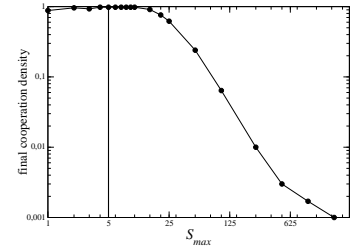}
\caption{Behaviour of the final frequency of cooperative actions as a function of $S_{max}$. The remaining parameters are the same of
reference~\cite{suz05}, the vertical line for $S_{max}=5$ specifies the value utilized in reference~\cite{suz05}. Each point averaged over 1000 realizations.}
\label{Smax_check}
\end{figure}

Finally, changing the value of $S_{max}$, we see that up to $S_{max}\simeq15$, the behaviour of the system is rather homogeneous, as shown
in Fig.~\ref{Smax_check}.

\ 

Our results show that the behaviour of the model is actually robust
for a large range of the parameters at stake, thus replicating Suzuky and Akiyama's results.

\ 

\subsection{Small-world networks}

\begin{figure}[tbp]
\centering
\includegraphics*[width=0.75\hsize]{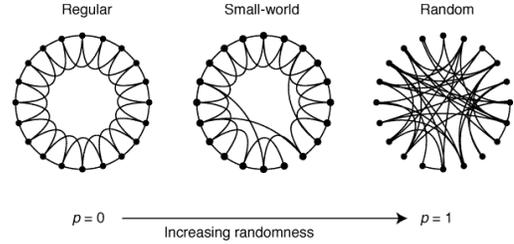}
\caption{Construction of SWN according Watts-Strogatz procedure. From reference~\cite{wat98}.}
\label{swn}
\end{figure}

In order to enlarge the scope of the model, we inserted network structure in it, thus introducing some adaptations into the original model. The first change we made was in the mechanism of assortment. In the original model, every player had the same probability to interact with every other agent, therefore the population is placed on a total connected graph (CG). This configuration is rather unrealistic, especially when we consider
groups bigger than a given size. It is then interesting to test the model behaviour over more realistic, even though still abstract, networks.
The first example we take under consideration is the so-called small-world network (SWN), as conceived by Watts and Strogatz in~\cite{wat98}.
In short, a SWN, is a regular ring with few short-cuts linking originally far away nodes. It is constructed as shown in Fig.~\ref{swn}: we start from
a ring where each node is connected with $2k$ nearest neighbours. Then, with probability $p$, each link is rewired (one of the node is left fixed,
the other is changed), so that it finally leads to the creation of a network with $pNk$ short-cuts. As shown in reference~\cite{wat98}, for $1/Nk<p<1/10$ the network
shows the typical small-world effect: even though at local level the system behaves as a regular lattice, i.e., an individual placed in a
SWN cannot distinguish the network from a regular one just watching his/her neighbours (high clustering coefficient), at a global level the
average distance between two randomly selected individuals is very low (proportional to the logarithm of the system size), unlike the regular
case.

In order to make the model work with this topology, we had to adapt the model dynamics to the specific situation. In particular, instead of extract $g$
agents at each round, we picked up a single player at each round and $g-1$ of its neighbours. In order to be sure that each individual had at least $g-1$
neighbours, we set $k=g-1$. Moreover, at the end of each generation, the offspring was randomly placed on the preexistent network, which is
defined at the beginning and does not change until the end of the simulation. Anyway, averaging over different realizations, each one has
its own networks, so that the averages are also over the topology.

\begin{figure}[tbp]
\centering
\includegraphics*[width=0.5\hsize]{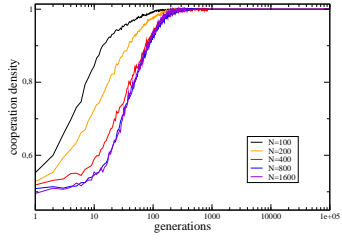}
\caption{Behaviour of the frequency of cooperative actions in a SWN with $p=0.05$ as a function of the number of generations for $g=2$
and different values of $N$ (from top to bottom: 100, 200, 400, 800 and 1600). The remaining parameters are the same of reference~\cite{suz05}.
Each curve averaged over more than 1000 realizations.}
\label{sw_check1}
\end{figure}

\begin{figure}[tbp]
\centering
\includegraphics*[width=0.5\hsize]{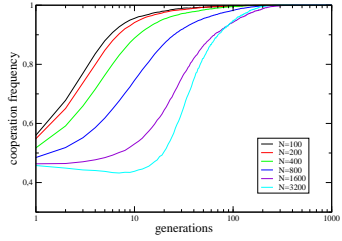}
\caption{Behaviour of the frequency of cooperative actions in a SWN with $p=0.05$ as a function of the number of generations for $g=4$
and different values of $N$ (100, 400 and 1600). The remaining parameters are the same of reference~\cite{suz05}. Each curve averaged
over more than 1000 realizations.}
\label{sw_check2}
\end{figure}

In Fig.~\ref{sw_check1} and~\ref{sw_check2} we see the cooperation frequencies for two values of $g$ and different sizes
of the system. Basically, in particular for $g>2$, the system shows an interesting behaviour: in particular, the dynamics is
always driving the system towards the achievement of complete cooperation, even though the timing can vary: full cooperation is reached
more rapidly for small values of $N$, whilst it can take up to thousands of generations for larger systems. It is worth to notice,
from Fig.~\ref{sw_check2}, that this consensus time seems to reach its limit value already for $N=3200$; in such case we can also distinguish
an initial small decrease of cooperation rate before the final (steep) increase.

In short, these results demonstrate that the small-world topology in itself makes full cooperation possible, although not so fastly as in different
configurations, as we are going to show in the next sections.

\ 

\subsection{Bipartite graphs}

\begin{figure}[tbp]
\centering
\includegraphics*[width=0.75\hsize]{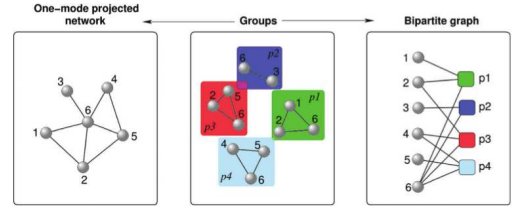}
\caption{Structure of a bipartite graph compared with a classical network. From reference~\cite{gar11}.}
\label{bg}
\end{figure}

Another topological configuration that accounts better for the complexity of real interactions among individuals is the so-called
bipartite graph (BG)~\cite{die97,gar11}. 
A bipartite representation contains two types of nodes denoting agents and groups, respectively. It implies that connections can be established only between nodes of different types and no direct connection among individuals is
allowed. Thus, such a bipartite representation preserves the information about the group structure: if two individuals
belong to the same three groups, they are ``more'' connected than two other individuals who are members of the same group. These two pairs would be equally represented in the classical one-mode projected network, while with the bipartite graph
this mesoscopic level of interactions is better depicted, as illustrated in Fig.~\ref{bg}.

Also in this case we adapted the original dynamics of the model to make it work on this kind of network. In particular,
the graph has $N$ individuals distributed into $M$ groups, each group composed of $g$ members. At the beginning of each round, the network is built
in this way: given $F\in(0,1)$, we set $gF$ initial members for each groups so that each individual belongs exclusively to one group. For
instance, if $N=150$, $g=20$ and $F=0.75$ (then $M=10$), at this stage we would have 15 agents in the first group, other
15 in the second one and so on until the last 15 in the tenth group. Then, each group must be completed choosing $(1-F)g=5$
individuals from the pool of those which do not belong to the group already. This can be accomplished in two different ways: first, by randomly picking $(1-F)g$ agents among the rest of the population; second, by selecting them according to their reputation, {\it i.e.}, their
image scores. When partner selection is available, an external player is randomly selected by the group, but accepted only if its image score is positive. Only if there is no
player in the whole population with good reputation, a candidate with negative image score is accepted in the group. Alternative ways of implementing partner selection were tested, like for example, accepting candidates with
image score equal or larger than the average strategy of the initial member of the group, but this did not produce any appreciable effects on the outcome of the simulations. Once the network is completely defined, each group plays a round of the game, with the same rules working
on CG and SWN. The procedure (network construction followed by a round of the game of each group) is repeated 10 times, then
the evolution process takes place again following the same rule given of the previous cases.

\begin{figure}[tbp]
\centering
\includegraphics*[width=0.5\hsize]{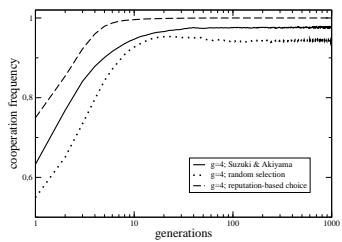}
\caption{Behaviour of the frequency of cooperative actions in a BG with as a function of the number of generations for $g=4,$
and $F=0.75$. The remaining parameters are the same of reference~\cite{suz05}. Each curve averaged
over 1000 realizations.}
\label{bg_check04}
\end{figure}

\begin{figure}[tbp]
\centering
\includegraphics*[width=0.5\hsize]{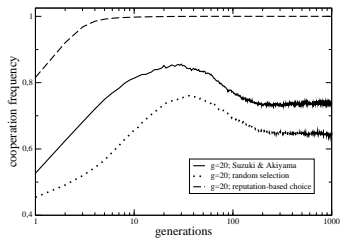}
\caption{Behaviour of the frequency of cooperative actions in a BG with as a function of the number of generations for $g=20$
and $F=0.75$. The remaining parameters are the same of reference~\cite{suz05}. Each curve averaged
over 1000 realizations.}
\label{bg_check20}
\end{figure}

In Fig.~\ref{bg_check04} and~\ref{bg_check20} we show the behaviour of the model for $N=200$ (or the closest integer compatible with the remaining parameters),
$F=0.75$, with the other parameters equal to the ones utilized by Suzuki and Akiyama~\cite{suz05}.

Our results show that the final cooperation level is {\it lower} here then in the CG
case if the added members of the groups are selected at random. However, when reputation-based partner selection is available in a population distributed on a bipartite graph,  full cooperation is achieved in a very short amount of time (about ten generations), and this is true also for large groups ($g=20$ in figure). This result does not depend on $F$: even when
partner selection is restricted to a small percentage of agents, it can favour the invasion of the cooperative strategies throughout the system. This effect can be explained by the fact
that, in general, in PGG it is better for individuals to get involved in as many groups as possible in order to maximize their
income~\cite{hau03}. However, if this is not linked to a reputation-based partner selection mechanism, defection is still very profitable and cooperators are driven out of the system. On the contrary, if reputation is used to select group members, having a positive image score has a positive effect on fitness. 

\ 

\subsection{Final strategy distributions}

In the model by Nowak and Sigmund~\cite{now98a,now98b}, based on the same image score mechanism, when the system ends up in a
final configuration of complete cooperation, the only surviving strategy is usually $k=0$, that is, the ``winning'' strategy is a
rather moderately generous one. A similar behaviour appears with our model in CG and SW topologies.

\begin{figure}[tbp]
\centering
\includegraphics*[width=0.5\hsize]{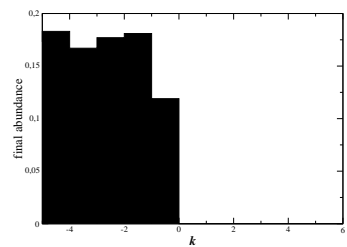}
\caption{Final (relative) strategy abundance for a system on BG, same system of Fig.~\ref{bg_check04} ($g=4$) with reputation-based choice
of the added members of each group. Values averaged over 1000 realizations.}
\label{DISTR_check}
\end{figure}

On the other hand, when working on BG topology, the final system configuration, always totally cooperative, presents all the negative
strategies, {\it i.e.} the more cooperative ones, as shown in Fig.~\ref{DISTR_check}. This means that taking account more carefully of the real
properties of the social interactions among idividuals not only enhances hugely cooperation to spread throughout the whole population,
but allows the survival also to the most generous and altruistic strategies.

\ 

\ 

\section{Discussion}
In a PGG in which the story of agents' past interactions is publicly available as an image score, cooperation can emerge and be maintained for small groups of agents. When we move from a mean-field situation to a small-world network, we observe that cooperation becomes stable after one hundred generations and for $g=4$. The real improvement is achieved thanks to the introduction of a partner choice mechanism on a bi-partite graph, where if a small percentage of group members are chosen on the basis of their reputations, cooperation can thrive. 

In a social dilemma the introduction of a reputation mechanism for partner selection on a bipartite graph makes deception unprofitable, thus cooperators can thrive. In such an environment, agents with a positive reputation are more socially desirable, thus they can enter several groups in which their contributions help to achieve the social optimum. On the other hand, defectors with negative reputations are actively avoided, thus driving them to complete extinction after ten generations. Even more striking is the fact that, unlike other models~\cite{suz05,bra03}, full cooperation is maintained even when group size increases. 

\ 

\ 

\section{Conclusions and perspectives}
The puzzle of the evolution of cooperation in humans can be successfully addressed if we take into account features of human societies that could have paved the way for the emergence of cooperative behaviors, like social networks and reputation. Moving from a replication of Suzuky and Akiyama~\cite{suz05} we showed that cooperation can emerge and be maintained in groups of agents playing a PGG on a network. We used two network topologies with different group and total population size, finding interesting differences especially in terms of the maximum level of cooperation achieved. Our results show that when partner selection is available in an affiliative network, cooperation can be easily reached even in large groups and for large system size.

The importance of social institutions~\cite{ost90} and informal social control~\cite{gia12,bes09} is well known to social scientists who, like Ellickson~\cite{ell91}, have stressed the importance of these features:
{\it ``A close-knit group has been defined as a social network whose members have credible and reciprocal prospects for the application of power against one another and a good supply of information on past and present internal events [\dots]. 
The hypothesis predicts that departures from conditions of reciprocal power, ready sanctioning opportunities, and adequate information are likely to impair the emergence of welfare-maximizing norms''} (p. 181).

Introducing a small world network does not alter the dynamics of cooperation in a PGG in a fundamental way, and this is also true for a bipartite graph with random partner selection. However, when we model the world as made of groups that can actively select at least one of their members, cooperators outperform free-riders in an easy and fast way. The evolutionary dynamics of our model can be linked to have a proximate explanation in psychological mechanisms for ostracism and social exclusion, two dreadful outcomes for human beings~\cite{abr05,bau95}. 
In large groups of unrelated individuals, direct observation is not
possible, and usually records of an individual's past behaviors are
not freely and publicly available. What is abundant and costless is
gossip, i.e., reported information about others' past actions, that
can be used to avoid free-riders, either by refusing to interact with them,
or joining another crew in which free-riders are supposedly absent. For this reason we plan to run simulations in which agents will be able to report private information about their past experiences, thus overcoming the unrealistic limitations posed by image score. We posit that the combination of a bi-partite graph social structure and gossip like exchanges will mimic human societies better and will provide useful insights about the evolution of cooperation in humans.

\ 

\ 

\section*{ACKNOWLEDGEMENTS}

\ 

We gratefully acknowledge support from PRISMA project, within the
Italian National Program for Research and Innovation (Programma Operativo
Nazionale Ricerca e Competitivit\'a 2007-2013. Settore: Smart Cities
and Communities and Social Innovation Asse e Obiettivo: Asse II -
Azioni integrate per lo sviluppo sostenibile).

\  

\

\end{document}